\documentclass[12pt,hyper]{JHEP3}  
\pdfoutput=1
\usepackage{amsmath}
\usepackage{amssymb}
\usepackage{graphicx}
\usepackage{multirow}
\usepackage{epsfig}
\usepackage{rotating}
\usepackage{longtable} 
\keywords{E6, Extra neutral gauge bosons}

\newcommand{\cc}{\c c}

\title{Alternative $Z'$  Bosons   in  $E_6$}
\author{Eduardo Rojas$^{1,2}$ and Jens Erler$^{3}$
\\
$^{1}$Instituto  de F\'isica,  
Universidad de Antioquia, \\Calle 70 No. 52-21, Medell\'in, Colombia. \\
$^{2}$Laboratorio de F\'isica 
Te\'orica e Computa\cc\~ao Cient\'ifica, 
Universidade Cruzeiro do Sul, 01506-000, S\~ao Paulo, Brazil. \\
$^3$ Departamento de F\'isica Te\'orica, Instituto de F\'isica, \\ 
Universidad Nacional Aut\'onoma de M\'exico, 04510 M\'exico D.F., M\'exico \\
\email{eduardo.rojas@cruzeirodosul.edu.br} \\
\email{erler@fisica.unam.mx}}

\newcommand{\be}{\begin{equation}}
\newcommand{\ee}{\end{equation}}
\newcommand{\bea}{\begin{eqnarray}}
\newcommand{\eea}{\end{eqnarray}}
\newcommand{\ba}{\begin{array}}
\newcommand{\ea}{\end{array}}

\newcommand{\ovl}{\overline} 
\newcommand{\ie}{\emph{i.e.,\ }}

\newcommand{\ubar}{\ovl{u}} 
\newcommand{\dbar}{\ovl{d}} 
\newcommand{\ddbar}{\ovl{D}} 
\newcommand{\nubar}{\ovl{\nu}} 
\newcommand{\lbar}{\ovl{L}} 
\newcommand{\eplus}{e^+}

\keywords{Extra neutral gauge bosons, $E_6$}

\date{}

\abstract{We classify the quantum numbers of the extra $U(1)'$ symmetries contained in $E_6$. 
In particular, we categorize the cases with rational charges and present the full list of models 
which arise from the  chains of the maximal subgroups of $E_6$. 
As an application, the classification allows us to determine all embeddings  
of the Standard Model fermions in all possible   decompositions of the fundamental  representation  of $E_6$ 
under its maximal subgroups. 
From this we find alternative chains of subgroups for Grand Unified Theories. 
We show how many of the known models including some new ones appear in alternative breaking patterns.
We also use low energy constraints coming from parity-violating asymmetry measurements and atomic parity non-conservation
to set limits on the $E_6$ motivated parameter space for a $Z'$ boson mass of~1.2~TeV.  
We include projected limits  for the present and upcoming QWEAK, MOLLER and SOLID experiments.
}
\preprint{}
\begin{document}

\section{Introduction}
\label{sec:intro}
Heavy neutral gauge bosons are a generic prediction of many types of new physics beyond the Standard Model~(SM). 
This is because extra $U(1)'$ symmetries serve as an important model-building tool 
(for example, to suppress phenomenologically strongly constrained processes)
giving rise --- after spontaneous $U(1)'$ symmetry breaking --- to physical $Z'$ vector bosons. 
Thus, with the advent of LHC proton-proton collisions at a center of mass energy of 13~TeV,
there exists a real possibility for the on-shell production of a $Z'$ boson~\cite{Langacker:2008yv,Salazar:2015gxa}. 

All representations of the $E_6$ gauge group~\cite{Gursey:1975ki,Achiman:1978vg} are anomaly-free and 
the fundamental {\bf 27}-dimensional representation is chiral and can accommodate a full SM fermion generation.
As a consequence, $E_6$-motivated $Z'$ bosons arise naturally in many popular extensions of the SM~\cite{Langacker:2008yv,Robinett:1982tq,London:1986dk},
both in top-down and bottom-up constructions.
Some of the $E_6$ subgroups, such as the original unification groups, $SU(5)$ and $SO(10)$,
and the gauge group of left-right models, $SU(4)\times SU(2)_L\times SU(2)_R$, play central roles in some of the best motivated extensions of the SM.   
Furthermore, the complete $E_6$-motivated $Z'$ family of models appears in a supersymmetric bottom-up approach 
exploiting a set of widely accepted theoretical and phenomenological requirements~\cite{Erler:2000wu}.
The one-parameter $Z'$ families~\cite{Carena:2004xs}, ${\bf 10} + x {\bf \bar{5}}$, $d - x u$ and $q + x u$,
where ${\bf 10}$ and ${\bf \bar{5}}$ are $SU(5)$ representations, $q$, $u$ and $d$ indicate $U(1)'$ quantum numbers proportional to
the SM quark doublets and singlets, and $x$ is an arbitrary real parameter, 
can also be discussed within the $E_6$ framework~\cite{Erler:2011ud}.

For all these reasons there is an expectation that an $E_6$ Yang-Mills theory,
or a subgroup of $E_6$ containing the SM in a non-trivial way,
might be part of a realistic theory~\cite{Slansky:1981yr}.
And if a heavy vector boson is seen at the LHC or at a future even more powerful collider,
aspects of the $E_6$ symmetry group will be central to the discussion of what this resonance might be telling us 
about the fundamental principles of nature.
 
However the discrimination between $Z'$ models could be challenging at the LHC due to the small number 
of high resolution channels at hadron colliders.
Another reason why the determination of the underlying symmetry structure is not straightforward 
is that the mass eigenstate of the $Z'$ is in general a linear combination of some of the underlying $Z'$ charges,
with the ordinary $Z$ boson of the SM mixed in.
Hence, it is useful to reduce the theoretical possibilities or at least to have a manageable setup.
This work represents an attempt in this direction and serves to spotlight a few tens of models in the two-dimensional   
space of $E_6$-motivated $Z'$ models.

All the $E_6$ breaking patterns and branching rules have been tabulated in Ref.~\cite{Slansky:1981yr}. 
The work by Robinett and Rosner~\cite{Robinett:1982tq} (hereafter referred to as RR)
showed several embeddings of the SM in the decomposition ${\bf 27} = {\bf(2,\bar{6})}+{\bf(1,15)}$ of the  fundamental representation of $E_6$
under $SU(2)\times SU(6)$.
Our aim here is to present an extended and more complete picture of this subject.
The first goal is to find alternative chains of subgroups for Grand Unified Theories~\cite{Barr:1981qv},
which will subsequently be a useful tool towards a systematization of $Z'$ bosons within the $E_6$ class. 

The paper is organized as follows:
In Section~\ref{sec:models} we review two different parameterizations for $Z'$ models based on the $E_6$ gauge group.
In Section~\ref{sec:class} we introduce a general classification of the $E_6$-motivated $Z'$ models with rational charges.
In Section~\ref{sec:breakings} we present all the $E_6$ chains of maximal subgroups involving  $U(1)$ symmetries and show the 
corresponding $Z'$ charges and their ($\alpha$, $\beta$) coordinates with respect to one of the parameterizations in Section~\ref{sec:models}.
Section~\ref{sec:low} shows the exclusion limits and reach for recent and upcoming low-energy experiments 
for the entire $E_6$-motivated $Z'$ parameter space for a $Z'$ boson mass of $M_{Z'}=1.2$~TeV. 
These low-energy measurements are competitive and highly complementary to both lepton and hadron colliders at the energy frontier.

It is important to remark that many interesting phenomenological models  appear 
in a natural way in $E_6$ breaking patterns, such as the leptophobic $Z_{\not{L}}$, 
$Z'$ bosons which at zero momentum transfer are proton-phobic, $Z_{\not{p}}$, or neutron-phobic, $Z_{\not{n}}$, 
$Z'$ bosons from supersymmetric models, as for example the $ Z_N$~\cite{Ma:1995xk,King:2005jy}, {\em etc}.  
Section~\ref{sec:breakings} (Table~\ref{tab:models1}) illustrates how the best known $Z'$ models arise naturally in this way.  

\section{The $E_6$ Parameterizations}
\label{sec:models}

Any three linearly independent $U(1)$ subgroups of $E_6$ can be used as a basis for the $Z'$ models within this group.
Once the normalization is fixed, the corresponding parameter space can be mapped to the surface of a three-dimensional sphere
which can be parameterized by two angles (the rank of $E_6$ exceeds that of the SM by two).
The $\alpha$ and $\beta$ parameters introduced in Refs~\cite{Erler:2011ud,Erler:1999nx} are the corresponding angles  
for the  orthonormal basis  $Z_\chi$,  $Z_\psi$ and  $Z_Y$,
\be
\label{eq:e6models}
Z' = \cos\alpha \cos\beta\, Z_\chi + \sin\alpha \cos\beta\,  Z_Y + \sin\beta\,  Z_\psi 
    = \frac{c_1\, Z_R + \sqrt{3}\, (c_2\, Z_{R_1} + c_3\, Z_{L_1})}{\sqrt{c_1^2 + 3\, (c_2^2 + c_3^2)}}\ .
\ee
Here, the $Z_Y$ refers to hypercharge, and the  $Z_\chi$ and $Z_\psi$ are defined through the breaking patterns
$SO(10) \to SU(5) \times U(1)_\chi$ and $E_6 \to SO(10) \times U(1)_\psi$, respectively.
For further details and charge assignments see~Refs.\cite{Langacker:2008yv,Robinett:1982tq}. 
The second form appearing in Eq.~(\ref{eq:e6models}) uses a different orthogonal basis~\cite{Erler:2011ud},
$U(1)_R$, $U(1)_{R_1}$, and $U(1)_{L_1}$, which are the maximal subgroups~\cite{Robinett:1982tq} 
defined by $SU(3)_{L,R} \to SU(2)_{L,R} \times U(1)_{L_1,R_1}$ and $SU(2)_R \to U(1)_R$, referring here 
to the trinification subgroup~\cite{Achiman:1978vg} of $E_6 \to SU(3)_C \times SU(3)_L \times SU(3)_R$.

In this parameterization the angles are replaced by the parameters $c_1$, $c_2$ and $c_3$, as indicated in Eq.~(\ref{eq:e6models}),
together with a normalization constraint. 
In general the $c_i$ are real numbers but in the most interesting cases we can usually choose them to be small integers 
by taking a convenient normalization. 
In Eq.~(\ref{eq:e6models}), $-\pi/2 < \beta \leq \pi/2$ is the mixing angle between the $U(1)_\chi$ and $U(1)_\psi$ charges, and 
$-\pi/2 < \alpha \leq \pi/2$ is non-vanishing when there is a mixing term~\cite{Holdom:1985ag} between hypercharge and the $U(1)'$.
Note, that any kinetic mixing term between the hypercharge and $U(1)'$ field strength tensors 
can be absorbed into the value of $\alpha$.

The $U(1)'$ charges of the particles appearing in the fundamental representation of $E_6$ are shown in 
Table~\ref{tab:e6charges} in terms of the parameters $c_1$, $c_2$ and $c_3$, satisfying
\begin{equation}
   \tan\alpha = \frac{c_1 + c_2 + c_3}{\sqrt{\frac{2}{ 3}}\, c_1 - \sqrt{\frac{3}{ 2}}\, (c_2 + c_3)}, \hspace{24pt}
   \tan\beta = \frac{{\rm sgn} [\frac{2}{ 3}\, c_1 - (c_2 + c_3)]}{\sqrt{\frac{2}{3}\, c_1^2 + (c_2 + c_3)^2}}  (c_3 - c_2).
\end{equation}

In the $E_6$ normalization for the hypercharge the electric charge is given by
\begin{align}
Q_{em}=T_3+\sqrt{\frac{5}{3}}Y, 
\end{align}
where $T_3$ is the third component of weak  isospin, which is   $1/2$ for the neutrino.  The hypercharge 
components in this normalization  are given by Eq.~(\ref{eq:e6models}) with $c_1 = 3$, $c_2 = 1$ and $c_3 = 1$.

\TABLE[t]{\label{tab:e6charges}
\begin{tabular}{|ccrr|ccrr|}
\hline
$l \equiv \left( \ba{c} \nu \\ e^- \ea \right)$ & & $-2 c_2$ & $-c_3$ &
$\ba{c} \bar{\nu} \\ e^+ \ea$ & $\ba{r} -c_1 \\ +c_1 \ea$ & $\ba{r} +c_2 \\ +c_2 \ea$ & $\ba{r} +2 c_3 \\ +2 c_3 \ea$ \\
\hline
$q \equiv \left( \ba{c} \phantom{l} u \phantom{l} \\ \phantom{l} d \phantom{l} \ea \right)$ & & & $+c_3$ &
$\ba{c} \bar{u} \\ \bar{d} \ea$ & $\ba{r} -c_1 \\ +c_1 \ea$ & $\ba{r} -c_2 \\ -c_2 \ea$ & \\
\hline\hline
$L \equiv \left( \ba{c} N \\ E^- \ea \right)$ & $-c_1$ & $+c_2$ & $-c_3$ &
$\ba{c} D \\ \ovl{D} \ea$ & & $\ba{r} \\ +2 c_2 \ea$ & $\ba{r} -2 c_3 \\ {} \ea$ \\
\hline
$\ovl{L} \equiv \left( \ba{c} E^+ \\ \ovl{N} \ea \right)$ & $+c_1$ & $+c_2$ & $-c_3$& 
$S$ & & $\ba{r} -2 c_2 \ea$ & $\ba{r} +2 c_3 \ea$ \\
\hline
\end{tabular} 
\caption{Charge assignment~\cite{Erler:2011ud} for the left-handed particles and antiparticles 
contained in a $\bf{27}$-dimensional representation of $E_6$
(the right-handed particles and antiparticles transforming in the antifundamental $\bf\overline{27}$ representation are implied).
The upper part of the table corresponds to the $\bf{16}$-dimensional representation of $SO(10)$, while the lower part shows the 
$\bf{10}$ (with an extra anti-quark weak singlet, $\ovl{D}$, of electric charge $-1/3$ and an additional weak doublet, $L$,
as well as their SM-mirror partners) and the~$\bf{1}$ (a SM singlet, $S$).  
This represents one fermion generation, and we assume family universality throughout.
The correct normalization (\ie the one which is directly comparable to the usual normalization of the gauge couplings of $SU(3)_C$
and $SU(2)_L$ of the SM) of these charges is obtained upon division by $2 \sqrt{c_1^2 + 3\, (c_2^2 + c_3^2)}$.}} 

\section{The  $E_6$  Structure}
\label{sec:class}

\subsection{Decomposition of the ${\bf 27}$ under $SU(2)\times SU(6)$}
The most important maximal subgroups of $E_6$ are $SO(10)\times U(1)$, $SU(6)\times SU(2)$, and $SU(3)\times SU(3)\times SU(3)$. 
The representation theory of compact Lie algebras~\cite{Slansky:1981yr} implies
that in the breaking $E_6\rightarrow SU(6)\times SU(2)$ the fermions in the ${\bf 27}$ are grouped into two multiplets,
${\bf 27} \to (\bf 2,\bf\bar{6})+(1,15)$.
The multiplet $(\bf 2,\bar{6})$ contains six $SU(2)$ doublets whereas fields in the $(\bf 1,15)$ multiplet are singlets under $SU(2)$.
There are four different ways to assign the SM fermions to a $\bf 27= (\bf 2,\bf\bar{6})+(1,15)$.
Namely, for $E_6\rightarrow SU(2)_X\times SU(6)$, where $X= L$ (left), $R$ (right), $I$ (inert), and $A$ (alternative), we have
\begin{align}
({\bf 2},{\bf \bar{6}})_L=&\left(L,\lbar,q,l\right)                        &({\bf 1},{\bf 15})_L=&(\nubar,S,\eplus,\dbar,\ubar,D,\ddbar)&SU(2)_L&\\
({\bf 2},{\bf \bar{6}})_R=&\left((\dbar,\ubar),(\lbar,L),(\eplus,\nubar)\right)  &({\bf 1},{\bf 15})_R=&(l,q,D,\ddbar,S)                     &SU(2)_R&\\
({\bf 2},{\bf \bar{6}})_I=&\left((\ddbar,\dbar),(L,l),(\nubar,S)\right)          &({\bf 1},{\bf 15})_I=&(\lbar,q,\ubar,D,\eplus)             &SU(2)_I&\\
({\bf 2},{\bf \bar{6}})_A=&\left((\ubar,\ddbar),(l,\lbar),(S,\eplus)\right)      &({\bf 1},{\bf 15})_A=&(L,q,\dbar,D,\nubar)                  &SU(2)_A&
\end{align}
RR~\cite{Robinett:1982tq} considered the cases $X= L,R,I$.
Here we add the embedding, $X=A$, to obtain a more symmetric and complete picture of the $E_6$ subgroups and models. 
The need of this embedding will become evident from the classification. 
{\rm
One can obtain the {\bf 15} representation from the tensor product  ${\bf 6} \times {\bf 6} = {\bf 21}_s + {\bf 15}_a$.
Specifically, for $SU (2)_A\times SU(6)$ we have explicitly (displaying only the upper-right parts of the matrix),
\begin{align} \nonumber
{\bf  15}=
\left[ \begin{pmatrix}
({\bf 3}_C,{\bf 1}_L)\\
({\bf 1}_C,{\bf 2}_L)\\
({\bf 1}_C,{\bf 1}_L)\\
\end{pmatrix}
\times
\begin{pmatrix}
({\bf 3}_C,{\bf 1}_L)&
({\bf 1}_C,{\bf 2}_L)&
({\bf 1}_C,{\bf 1}_L)
\end{pmatrix} \right]_a=
\\ 
\left(
\begin{array}{c|c|c}
(\bar{{\bf 3}}_C,{\bf 1}_L)&({\bf 3}_C,{\bf 2}_L)    &({\bf 3}_C,{\bf 1}_L)\\
\hline
    &({\bf 1}_C,{\bf 1}_L)    &({\bf 1}_C,{\bf 2}_L)  \\
\hline  
       &             & 0
\end{array}
\right)
=
\left(
 \begin{array}{ccc|cc|c}
   0 & \dbar_3 & \dbar_2 & d_1 & u_1 & D_1\\
     &  0      & \dbar_1 & d_2 & u_2 & D_2\\
     &         &       0 & d_3 & u_3 & D_3\\
\hline     
     &         &         & 0 &\nubar & E^{-}\\            
     &         &         &  & 0& N \\
\hline     
     &         &         &  &  & 0 
\end{array}            
\right).
\end{align}
The fermions in the anti-fundamental representation of $SU(6)$ are
${\bf \bar{6}}_{1/2}=\left(\ubar, l,S\right)$ and
${\bf \bar{6}}_{-1/2}=\left(\ddbar,\lbar,\eplus\right)$,
where the subscripts $\pm 1/2$ are the $U(1)_A$ charges. 
} 

\subsection{Alternative models}
We say that two $Z'$ models have the same multiplet structure if they can be obtained from one another 
by swapping some fermions between the multiplets.
In other words, they have equal numbers of multiplets and for every multiplet in one model 
there is the corresponding multiplet in the other model
with the same dimension and the same charges. 
For example, $U(1)_R$, $U(1)_I$ and $U(1)_A$ in Table~\ref{tab:models11} of Section~\ref{sec:breakings} have the same multiplet structure.
We refer to models with the same multiplet structure as a given $Z'$ as alternative models of this $Z'$. 

\subsection{Generalized RR notation} 
To begin with, we introduce a notation for the $Z'$ models based on the subgroups involved in specific breaking chains.
This notation borrows some elements of the work by RR~\cite{Robinett:1982tq}
and will be very useful to list subgroup chains:
\begin{equation}
U(1)\equiv
\begin{cases}
U(1)_{n-m\ m\ Z} & \text{for the $U(1)$ in }  SU(n)\rightarrow SU(n-m)\times SU(m)\times U(1)\\
U(1)_{n-1\ 1\ Z} & \text{for the $U(1)$ in }  SU(n)\rightarrow SU(n-1)\times U(1)\\
U(1)_{X} & \text{for the $U(1)$ in }  SU(2)_X\rightarrow U(1)
\end{cases}
\end{equation}
The subscript $Z= X,XY,\ovl{X}$ (the notation is introduced below) with $X,Y = R,I,A$, depends on the multiplets involved in the breaking,
and in the following we will often use the abbreviation $U_{n-m\ m\ Z}$ for $U(1)_{n-m\ m\ Z}$.
The $U_{R}$~\cite{Robinett:1982tq} does not couple to the left-handed projection of the SM fermions,  
the $U_I$~\cite{Robinett:1982tq} corresponds to the inert model which does not couple to up-type quarks, 
and similarly the $U_{A}$ does not couple to down-type quarks~\cite{Erler:2011ud}.  
The model   $U_{n-m\ m\ X}$, with  subscript $X$,  indicates that the charges are perpendicular to the $U(1)_X$,
\ie if $U_{n-m\ m\ X}(f)$ is the charge of fermion $f$ under $U_{n-m\ m\ X}$, 
then $\sum_{\text{f$\in \bf 27$ }}U_{n-m\ m\ X}(f) U_{X}(f)= 0$.
For the $U_X$ itself we have the normalization condition $\sum_{\text{f$\in \bf 27$ }}U_{X}^2(f)= 3$ 
(the eigenvalue of the quadratic Casimir operator for the $\bf 27$ in the standard normalization).
We use the notation  $U_{n-m\ m\ \ovl{X}}$,  for the  alternative model of $U_{n-m\ m\ X}$  which is also perpendicular to $U(1)_X$.  
There are two sets of  models labeled  with  $XY$,        
the alternative models of $U(1)_{\chi} \in SO(10)$, which are  referred to as $U(1)_{\chi XY}$ with $X\ne Y$.
These models  are perpendicular to $U(1)_{42X}$ and to $U(1)_{32Y}$. 
In a similar way we  define the models $U(1)_{41XY}$ which are defined to be perpendicular to $U_{51\ovl{X}}$ and to  $U_{31Y}$.
It is important to distinguish between the $Y$ used for hypercharge, 
and $Y= R,I,A$ which appears in the generalized RR notation and in the subscripts 
of the charges in Table~\ref{tab:ci}. 
 
A special case in RR is the model $U(1)_{33}$~(RR use the alternative notation $U(1)_{L1}$)
motivated by the breaking $SU(6)\rightarrow SU(3)_C\times SU(3) \times U(1)_{33}$.
Since a given $U(1)$ could appear in different breaking chains, there may be several notations for a single model.
{\em E.g.\/}, in the breaking $SU(3)_L \rightarrow SU(2)_L  \times  U(1)_{21L}$,  
the group $ U(1)_{21L}$ corresponds to $U(1)_{33}$; 
for that reason the alternative models of $U_{33}$ orthogonal to 
$U_{R}$, $U_{I}$, $U_{A}$ are $U_{21\bar{R}}$, $U_{21\bar{I}}$ and $U_{21\bar{A}}$, respectively,
as is shown in Table~\ref{tab:models10}. Because the $U_{33}$ is orthogonal 
to $U_{R}$, $U_{I}$ and $U_{A}$ we do not use the subscript $X$ as in other models.

\subsection{$U(1)'$ classification}
We will make use of the $SU(2)_X$ symmetries in order to implement a classification that identifies $Z'$ models with similar multiplet structures. 
For this we define $Q_{mnX}^{0}$ as the models with integer charges~(up to a normalization) in Eq.~(\ref{eq:e6models}) 
perpendicular to $U(1)_X$, \ie $\sum_{\text{f$\in \bf 27$ }}Q_{mnX}^{0}(f)U_{X}(f)= 0$, where $m,n$ are integers. 
The explicit forms of $Q_{nmX}^{0}$ are shown in Table~\ref{tab:ci}.
The most general form for a model that is  not  perpendicular to $U_X$ is the linear combination $c_1 Q_{00X}^1+Q_{nmX}^{0}$, 
with $c_1$ an integer different from zero and $Q_{00X}^1$ the charges of $U_X$. We label it as  $Q^l_{mnX}$ and the explicit form 
of the charges are shown in Tables~\ref{tab:ci} and \ref{tab:models3}.	 

In Table~\ref{tab:models3} we define $Q_{mnX}\equiv Q_{mnX}^{0}$, and $Q_{mnX}^{-l}$ as the conjugate of $Q_{mnX}^{l}$. 
For fixed $X$ the set $\{Q_{mnX}^l\}$ (with ${m,n,l\in \mathbb{Z}}$) covers all $E_6$ $Z'$ models, 
so that the $U(1)'$ charges of a model can be written in different bases as $Q_{mnX}^l$ and $Q_{m'n'Y}^{l'}$, 
with $X,m,n,l\ne Y,m',n',l'$. 
We choose as the systematic name of the model the one which minimizes  $\lvert l\rvert$ in such a way that $m,n,l$ are integers.  
For this convention the systematic name is uniquely defined in most of the cases.
In cases of ambiguity, it is always possible to apply a symmetry argument to arrive at a systematic nomenclature.
For example, if for a given model $\lvert l\rvert$ is a minimum for both $X=I$ and $X=A$ then we choose the unique name 
$Q_{m'n'R}^{l'}$, as is the case for the $U_R$, $U_A$ and $U_I$ models.

\TABLE[ht]{\label{tab:ci}
\noindent\makebox[\textwidth]{
\scalebox{0.9}{
\begin{tabular}{|l|rrr|}
\hline
            & $c_1$          &$c_2$          & $c_3$     \\
\hline
$Q_{mnR}^{l}$ & $l$         & $+n$       & $+m$  \\
$Q_{mnI}^{l}$ & $-(3n+l)/2$ & $-(n-l)/2$ & $+m$  \\
$Q_{mnA}^{l}$ & $(3n-l)/2$  & $-(n+l)/2$ & $+m$  \\
\hline
\end{tabular}}}
\caption[]{Coefficients of the  $Q_{mnX}^{l}$ charges in the $Z_R$, $Z_{R1}$, $Z_{L1}$ basis. 
The $Q_{mnX}^{0}$ are defined  as the models with integer charges perpendicular 
to $U_X$, \ie $\sum_{\text{f$\in \bf 27$ }}Q_{mnX}^{0}(f)U_{X}(f)= 0$.
Every set of  $Q_{mnX}^{l}$ with fixed $X$ contains all the $Z'$s in $E_6$ ($X= R,I,A$).}}

\TABLE[ht]{\label{tab:models3}
\noindent\makebox[\textwidth]{
\scalebox{0.7}{
\begin{tabular}{|l|l|l|l|l|l|l|l|l|l|l|l|l|l|}
\hline
$Q_{mnR}^{l}$  & $q_{m} $&&$D_{-2m} $&$\dbar_{l-n}$&$\ubar_{-l-n}  $&$\lbar_{l+n-m}$&$L_{-l+n-m}     $&$\eplus_{l+n+2m} $&$\nubar_{-l+n+2m}$&$\ddbar_{2n} $&$l_{-m-2n}     $&&$S_{2m-2n}$ \\
$Q_{mnI}^{l}$  & $q_{m} $&&$D_{-2m} $&$\ddbar_{l-n}$&$\dbar_{-l-n} $&$L_{l+n-m}    $&$l_{-l+n-m}     $&$\nubar_{l+n+2m} $&$ S_{-l+n+2m}    $&$ \ubar_{2n} $&$\lbar_{-m-2n} $&&$\eplus_{2m-2n}$ \\
$Q_{mnA}^{l}$  & $q_{m} $&&$D_{-2m} $&$\ubar_{l-n}$&$\ddbar_{-l-n} $&$l_{l+n-m}    $&$\lbar_{-l+n-m} $&$S_{l+n+2m}      $&$\eplus_{-l+n+2m}$&$\dbar_{2n}  $&$L_{-m-2n}     $&&$\nubar_{2m-2n}$ \\
\hline
$Q_{mnR}^{-l}$  & $q_{m} $&&$D_{-2m} $&$\ubar_{l-n} $&$\dbar_{-l-n}$&$L_{l+n-m}     $&$\lbar_{-l+n-m}$&$\nubar_{l+n+2m} $&$\eplus_{-l+n+2m} $&$\ddbar_{2n}   $&$l_{-m-2n}     $&&$S_{2m-2n}$ \\
$Q_{mnI}^{-l}$  & $q_{m} $&&$D_{-2m} $&$\dbar_{l-n} $&$\ddbar_{-l-n}$&$l_{l+n-m}    $&$L_{-l+n-m}    $&$S_{l+n+2m}      $&$ \nubar_{-l+n+2m}$&$   \ubar_{2n} $&$\lbar_{-m-2n} $&&$\eplus_{2m-2n}$ \\
$Q_{mnA}^{-l}$  & $q_{m} $&&$D_{-2m} $&$\ddbar_{l-n}$&$\ubar_{-l-n}$&$\lbar_{l+n-m} $&$l_{-l+n-m}    $&$\eplus_{l+n+2m} $&$S_{-l+n+2m}      $&$ \dbar_{2n}   $&$L_{-m-2n}     $&&$\nubar_{2m-2n}$ \\
\hline
$-Q_{nmR}^{l}$  & $\dbar_{-l+m}$&$\ubar_{l+m}  $&$\ddbar_{-2m}$&$q_{-n} $&$ $&$L_{l+n-m}     $&$\lbar_{-l+n-m}$&$l_{n+2m}        $&&$D_{2n}$&$ \eplus_{-l-m-2n}$&$\nubar_{l-m-2n} $&$ S_{2m-2n}$ \\
$-Q_{nmI}^{l}$  & $\ddbar_{-l+m}$&$\dbar_{l+m} $&$\ubar_{-2m} $&$q_{-n} $&$ $&$l_{l+n-m}     $&$ L_{-l+n-m}   $&$ \lbar_{n+2m}   $&&$D_{2n}$&$\nubar_{-l-m-2n} $&$ S_{l-m-2n}     $&$  \eplus_{2m-2n}$ \\
$-Q_{nmA}^{l}$  & $\ubar_{-l+m}$&$\ddbar_{l+m} $&$\dbar_{-2m}$&$q_{-n} $&$  $&$\lbar_{l+n-m} $&$l_{-l+n-m}    $&$ L_{n+2m}       $&&$D_{2n}$&$S_{-l-m-2n}      $&$\eplus_{l-m-2n} $&$ \nubar_{2m-2n}$ \\
\hline
\end{tabular}}}
\caption[]{
Fermion charge assignment  for the $E_6$-motivated $Z'$ models. $l=0$ corresponds to  set of models
with explicit $SU(2)_R$, $SU(2)_A$, or $SU(2)_I$ symmetry, \ie $Q_{mnX}^{0}\perp Q_{00X}^{1}$.
The alternative  models of $Q_{mnX}^{0}$ are  $Q_{nmX}^{0}$ with $X= R,I,A$ and $Q_{nmY}$ with $Y\ne X$, 
this becomes clear by comparing the top panel against  the bottom panel.
Further models can be obtained from these by splitting the $SU(2)$ doublets 
by adding (in general $l$ times) the $c_i$ of the corresponding $\pm Q_{X00}^1$.  
These are denoted by $Q^{l}_{mnX}$. 
For $l\ne 0$ the alternative models of $Q_{mnX}^l$ are $Q_{mnY}^{-l}$ where $Y$ may in general be different from $X$.
This becomes clear when one compares the top panel against the middle one. 
}}

\subsection{Alternative models in $E_6$}
\label{subsec:alternative}
As can be seen from the middle panel in Table~\ref{tab:models3}, 
for $l\ne 0$ the alternative models of $Q_{mnX}^l$  are $Q_{mnX}^{-l}$ with $X=R,I,A$ and $Q_{mnY}^{l}$ with $Y\ne X$.
For $l=0$,  $Q_{mnX}$ is self-conjugate, so in this case the  alternative  models of $Q_{mnX}$ are
$Q_{nmX}$ with $X= R,I,A$ and $Q_{nmY}$ with $Y\ne X$~(see the bottom panel in Table~\ref{tab:models3}).
In the generalized RR notation if  $Q_{mnX}= U(1)_{m'n'X}$ then $Q_{nmX}=U(1)_{m'n'\ovl{X}}$.
To summarize, we have
\TABLE[t]{\label{tab:models1}
\noindent\makebox[\textwidth]{
\scalebox{0.9}{
\begin{tabular}{|c|c|c|c|c|c|c|}
\hline
$Z'$                                     &
$ Z_R$~\cite{Robinett:1982tq}            & 
$ Z_{\not d}$~\cite{Erler:2011ud}        & 
$   -  Z_I$~\cite{Robinett:1982tq}       & 
$   -  Z_{L_1}$~\cite{Robinett:1982tq}   & 
$   -  Z_{R_1}$~\cite{Robinett:1982tq}   & 
$      Z_{\not p}$~\cite{Erler:2011ud}      \\ 
\hline
RR                & 
$ U_R$            & 
$ U_{A}$          & 
$ U_I $           & 
$ U_{33}$         & 
$ U_{21\ovl{R}}$  & 
$ U_{21\ovl{A}}$  
\\ 
$Q_{nm}^{l}$      &
$ Q_{R00}^1$      & 
$ Q_{A00}^1$      & 
$-Q_{I00}^1$      & 
$ Q_{X-10}$       & 
$ Q_{R0-1}$       &
$ Q_{A01}$   \\
\hline 
\hline     
$Z'$&  
$   -  Z_{\not n}$~\cite{Sanchez:2001ua,Erler:2011ud}& 
$   -  Z_{B-L}$~\cite{Pati:1974yy}   &  
$ Z_{ALR}$~\cite{Ma:1986we}                  & 
$-  Z_{\not L}$~\cite{Babu:1996vt}           & 
$ Z_\psi$~\cite{Robinett:1982tq}             & 
$ Z_\chi$~\cite{Robinett:1982tq}              
\\    
\hline
RR                  & 
$ U_{21\ovl{I}}$    & 
$ U_{31R}$          &
$ U_{31A}$          & 
$ U_{31I}$          & 
$ U_{42R}$          & 
$ U_{\chi RI}$       
\\
$Q_{mn}^l$          &
$ Q_{I0-1}$         & 
$ Q_{R-1-1}$        &
$ Q_{A11}$          & 
$ Q_{I-1-1}$        & 
$ Q_{R1-1}$         & 
$ Q_{A-23}^1$       
\\ 
\hline 
\hline     
$Z'$                                         &  
$ Z_N$~\cite{Ma:1995xk,King:2005jy}          &
$Z_{\chi^{*}}{[\text{flipped}-SU(5)]}$\cite{Barr:1981qv}&
$ Z_\eta$~\cite{Witten:1985xc}               & 
$ Z_Y$~\cite{Glashow:1961tr,Weinberg:1967tq} & 
$Z_S$~\cite{Erler:2002pr,Kang:2004pp}        &
---          
\\    
\hline
RR                  & 
$ U_{\chi AI}$      & 
$ U_{\chi RA}$      &
$ U_{51I}$          & 
$  U_{32I}$         & 
---                 &
---
\\
$Q_{mn}^l$          &
$-Q_{R-23}^{-1}$    &
$-Q_{I-23}^{-1}$    &
$ Q_{I-2-1}$        & 
$Q_{I1-2}$          & 
$Q_{A-14}^{3}$      &
--- 
\\    
\hline 
\end{tabular}}}
\caption{Systematic notation  $Q_{mn}^l$ and generalized RR notation for various $E_6$-motivated $Z'$ bosons. 
All of them appear in the literature.
The $Z_{\not p}$ and the $Z_{\not n}$ are bosons which do not couple --- at vanishing momentum transfer and at the tree level ---
to protons and neutrons, respectively.
Similarly, the $Z_{\not L}$, $Z_I$, and $Z_{\not d}$ bosons are blind, respectively, 
to SM leptons, up-type quarks, and down-type quarks.
The $Z_{B-L}$ couples purely vector-like while the $Z_\psi$ has only axial-vector couplings to the ordinary fermions. 
For $Q_{mn}^l$ we take the sign of the $\alpha$-$\beta$ parameterization Eq.~(\ref{eq:e6models}).
For convenience the models with the same multiplet structure of the  $Z_{\chi}$ are referred to as $U_{\chi XY}$.}}

\begin{align}\label{def:alt}
\text{alternative models of } Q^{l}_{X mn} =
\begin{cases}
\phantom{-} Q_{Y mn}^{-l} \text{ for any } Y=R,I,A  &\text{if\ } l\ne 0 \\  
\phantom{-}  Q_{Y mn}^{0}  \text{ for any}\ Y\ne X &\text{if\ } l = 0 \\
\phantom{-}  Q_{X nm}^{0} \ \ \ n \leftrightarrow m &\text{if\ } l= 0    \\
-Q^{l}_{X mn}
\end{cases}
\end{align}
After fixing the normalization, a global sign is still undefined. 
Indeed, reversing the overall sign in the charges leads, in principle, to a different model.  
While this sign is physical, we can absorb it in the $Z\text{-}Z'$ mixing angle, whose sign is then meaningful.
From now on, let us just consider models of the form~(\ref{eq:e6models}), \ie
without a global minus in front, 
$$Z' = \cos\alpha \cos\beta\, Z_\chi + \sin\alpha \cos\beta\,  Z_Y + \sin\beta\,  Z_\psi.$$

Since we are limiting the global sign to be positive 
the maximum number of  models with the same structure  in Eq.~(\ref{def:alt}) reduces from 12 to 6.
The above analysis is summarized in Eq.~(\ref{def:alt}) and is a way to show the implications of Table~\ref{tab:models3}.
Table~\ref{tab:models3} shows why our classification is useful and it constitutes  an important summary of the present work;
it is worth to notice that this table  is valid for any $Z'$ with rational charges in $E_6$. 
In Tables~\ref{tab:models10} and \ref{tab:models11} (see Section~\ref{sec:breakings}) 
we will make use of the property $Q_{-m-n}^{-l}=-Q_{mn}^{l}$
to write the charges in a way that better reflects the underlying structure.

\subsection{A geometrical interpretation}
\label{sec:geometrical}
\FIGURE[t]{\label{fig:E6}
\includegraphics[scale=0.5]{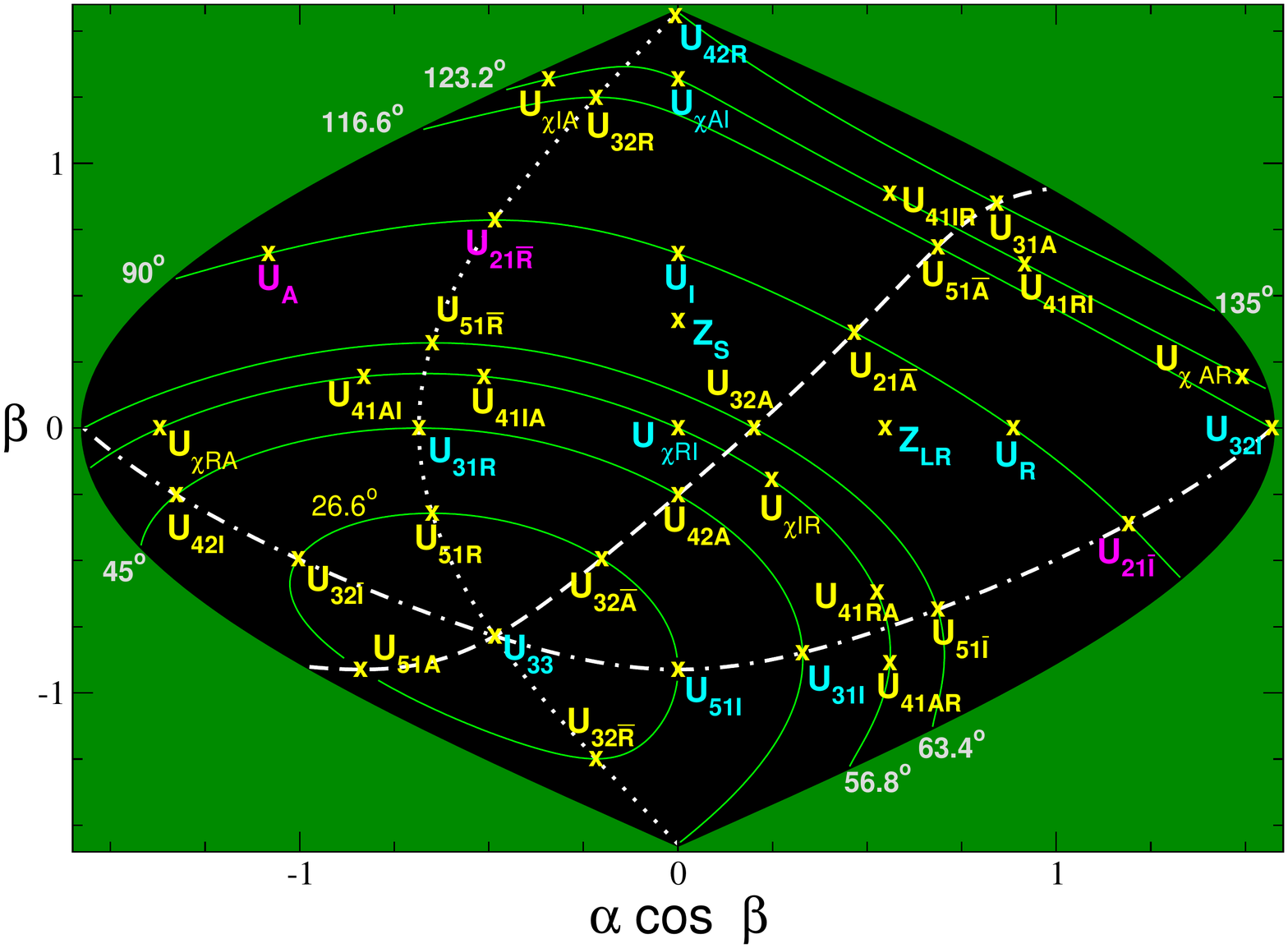}
\caption{$\alpha$-$\beta$ Sanson-Flamsteed projection of $E_{6}$ $Z'$ models.
The continuous green lines correspond to all  models  at a fixed angle of  $ U_{33}= -Z_{L1}$.  
The white dotted, dot-dashed and dashed lines correspond to the family of models 
perpendicular to $U_R$, $U_I$ and $U_A$, respectively. 
See the text for details. 
Labels in cyan correspond to very well known models in the literature
(for the conventional names see Table~\ref{tab:models1}).
Models with magenta labels are discussed in~\cite{Erler:2011ud}; the remaining models are indicated in yellow.
For every $U(1)$ it is possible to associate a three-dimensional vector in the $E_6$ parameter space, the angles in degrees correspond to the angle
with respect to $U(1)_{33}$ as explained in Section~\ref{sec:geometrical}.}}

A $U(1)'$ in $E_6$ can be written as a linear combination of an orthogonal vector basis as in Eq.~(\ref{eq:e6models}).
We can define the dot product between two models as $\sum_{\text{f$\in \bf 27$ }} Q_{mnX}^l(f) \  Q_{m'n'}^{l'}(f)$.
For a given $Z'$ (for $l\ne 0$) the modulus of the cosine of the angle between the models $\pm Q^{l}_{mnX}$ and  
$U_{33}$ is the same as the corresponding value between any of its alternative models and $U_{33}$. 
However there are two possible different signs for the cosine of the angle, 
namely $\mp \sqrt{3}m/(\sqrt{l^2+3(n^2+m^2)})$~(which are independent of  $X$).
Every sign corresponds to a curve in the $\alpha$-$\beta$ plane.  
In general, models with the same multiplet structure will appear on two different green continuous lines in Figure~\ref{fig:E6}.
In the case $l=0$ the modulus of the angle between models with the same multiplet structure and $U_{33}$ could be different
and, as in the case $l\ne 0$, the global sign of the model is also relevant.
Similarly, all the alternative models of a given set of charges $Q_{mnX}$ appear at most on two continuous green curves.  
All the models perpendicular to a fixed  $X$   are in a plane which contains the polar axis, which is generated by the vector $U_{33}$. 
The intersections of these planes, one for every $X$, with the surface of the sphere parameterized by $\alpha$ and $\beta$ 
are shown in Figure~\ref{fig:E6}
and correspond to the models perpendicular to $U(1)_R$~(dotted), $U(1)_I$~(dot dashed) and $U(1)_A$~(dashed). 
This geometrical interpretation gives us insight into the underlying structure, \ie 
under the present classification  the models with the same multiplet structure appear in a symmetric way around the pole,
which corresponds to the  $U_{33}$ model.

\TABLE{\label{tab:models10}
\noindent\makebox[\textwidth]{
\scalebox{0.88}{
\begin{tabular}{|l|l|l|rrr|cc|}
\hline
$U(1)^\prime$ & $Q_{mn}^{l}$ & & $c_1$ & $c_2$ & $c_3$&$\tan\alpha$&$\tan\beta $ \\ \hline
$U(1)_{R}$    & $+Q_{R00}^1$ & $+Z_{R}$ & $1$ & $0$ &$0$&$\sqrt{3/2}$&$0$ \\
$U(1)_{I}$    & $-Q_{I00}^1$ & $-Z_{I}$& $1$ & $-1$ &0&$0$&$\sqrt{3/5}$ \\
$U(1)_{A}$    & $+Q_{A00}^1$ &$+Z_{\not{d}}$& $-1$  & $-1 $   &$0$&$-2\sqrt{6}$&$\sqrt{3/5}$ \\ \hline
$U(1)_{33}$        & $+Q_{X-10}$ & $-Z_{L1}$ & $0$ & $0$ &$-1$&$-\sqrt{2/3}$&$-1$ \\
$U(1)_{21\ovl{R}}$ & $+Q_{R0-1}$ & $-Z_{R1}$ &$0$ & $-1$ &$0$&$-\sqrt{2/3}$&$1$ \\
$U(1)_{21\ovl{I}}$ & $+Q_{I0-1}$ & $-Z_{\not{n}}$ &$+3$ & $+1$ &$0$&$4\sqrt{2/3}$&$-1/\sqrt{7}$ \\
$U(1)_{21\ovl{A}}$ & $-Q_{A0-1}$ & $+Z_{\not{p}}$ &$+3$ & $-1$ &$0$&$2\sqrt{2/3}/3$&$1/\sqrt{7}$ \\ \hline
$U(1)_{31R}$       & $-Q_{R11}$ & $-Z_{B-L}$  &$0$ & $-1$ &$-1$&$-\sqrt{2/3}$&$0$\\
$U(1)_{31I}$       & $-Q_{I11}$ & $-Z_{\not{L}}$& $3$ & $1$ &$-2$&$2\sqrt{2/3}/3$&$-3/\sqrt{7}$ \\
$U(1)_{31A}$       & $+Q_{A11}$ & $+Z_{ALR}$     & $+3$ & $-1$ &$+2$&$4\sqrt{2/3}$&$3/\sqrt{7}$ \\ \hline
$U(1)_{42R}$       & $+Q_{R1-1}$ & $+Z_{\psi}$   & $0$ & $-1$ &$+1$&$0$&$\infty$ \\
$U(1)_{42I}$       & $-Q_{I1-1}$ &    ---      & $-3$ & $-1$ &$-2$&$-2\sqrt{6}$&$-1/\sqrt{15}$ \\
$U(1)_{42A}$       & $-Q_{A1-1}$ &    ---      & $+3$ & $-1$ &$-2$&$0$&$-1/\sqrt{15}$ \\ \hline
$U(1)_{32R}$       & $+Q_{R1-2}$ &    ---      & $0$ & $-2$ &$1$&$-\sqrt{2/3}$&$3$ \\
$U(1)_{32I}$       & $+Q_{I1-2}$ & $+Z_{Y}$     & $+3$ & $+1$ &$+1$&$\infty$&$0$\\
$U(1)_{32A}$       & $-Q_{A1-2}$ &  ---         &$+3$ & $-1$ &$-1$&$1/\sqrt{24}$&$0$\\
$U(1)_{32\ovl{R}}$ & $+Q_{R-21}$ &  ---        & $0$ & $1$ &$-2$&$-\sqrt{2/3}$&$-3$\\
$U(1)_{32\ovl{I}}$ & $+Q_{I-21}$ & ---         & $-3$ & $-1$ &$-4$&$-8\sqrt{2/3}/3$&$-3/\sqrt{31}$ \\
$U(1)_{32\ovl{A}}$ & $+Q_{A-21}$ & ---         &$+3$  & $-1$    &$-4$    &$-2\sqrt{2/3}/7$&$-3/\sqrt{31}$ \\ \hline
$U(1)_{51R}$       & $-Q_{R21}$ & ---          & $0$ & $-1$ &$-2$&$-\sqrt{2/3}$&$-1/3$\\
$U(1)_{51I}$       & $-Q_{I21}$ & $+Z_{\eta}$  & $+3$ & $+1$ &$-4$&$0$&$-\sqrt{5/3}$ \\
$U(1)_{51A}$      & $-Q_{A21}$  & ---          & $-3$ & $+1$ &$-4$&$-2\sqrt{6}$&$-\sqrt{5/3}$\\
$U(1)_{51\ovl{R}}$ & $-Q_{R12}$ & ---          &$0$ & $-2$ &$-1$&$-\sqrt{2/3}$&$1/3$\\
$U(1)_{51\ovl{I}}$ & $-Q_{I12}$ & ---          &$3$ & $1$ &$-1$&$\sqrt{3/2}$&$-\sqrt{2/3}$ \\
$U(1)_{51\ovl{A}}$ & $+Q_{A12}$ & ---          & $+3 $ & $-1 $ & $+1$&$\sqrt{3/2}$&$\sqrt{2/3}$ \\ \hline
$U(1)_{41IA}$& $+Q_{R-2-3}^{1}$ &  ---         & $+1$ & $-3$ &$-2$&$-4\sqrt{6}/17$&$\sqrt{3/77}$\\
$U(1)_{41AR}$ &$+Q_{I-2-3}^{1}$ &  ---         & $2$ & $1$ &$-1$&$\sqrt{3/2}$&$-\sqrt{3/2}$ \\
$U(1)_{41RI}$ & $-Q_{A-2-3}^{1}$&  ---         &$5$ & $-1$ &$+2$&$6\sqrt{6}/7$&$3\sqrt{3/53}$ \\
$U(1)_{41AI}$ &$+Q_{R-2-3}^{-1}$& ---          &$-1$ & $-3$ &$-2$&$-6\sqrt{6}/13$&$\sqrt{3/77}$\\
$U(1)_{41RA}$ &$+Q_{I-2-3}^{-1}$&  ---         &$5$ & $1$ &$-2$&$4\sqrt{6}/13$&$-3\sqrt{3/53}$\\
$U(1)_{41IR}$ &$+Q_{A-2-3}^{-1}$& ---          & $2$ & $-1$ &$+1$&$\sqrt{3/2}$&$\sqrt{3/2}$ \\ \hline
$U(1)_{\chi RI}$&$+Q_{A-23}^{1}$&$+Z_{\chi}$    & $2$ & $-1$ &$-1$&$0$&$0$\\
$U(1)_{\chi AR}$&$-Q_{I-23}^{1}$&---           & $5$ & $1$ &$2$&$8\sqrt{6}$&$\sqrt{3/77}$\\
$U(1)_{\chi IA}$&$-Q_{R-23}^{1}$&---           &$-1$ & $-3$ &$2$&$-2\sqrt{6}$&$\sqrt{15}$\\
$U(1)_{\chi IR}$&$+Q_{A-23}^{-1}$&---          & $5$ & $-1$ &$-2$&$2\sqrt{6}/19$&$-\sqrt{3/77}$\\
$U(1)_{\chi RA}$&$-Q_{I-23}^{-1}$&$+Z_{\chi*}{[\text{flipped}-SU(5)]}$  & $-2$ & $-1$ &$-1$&$-2\sqrt{6}$&$0$\\
$U(1)_{\chi AI}$&$-Q_{R-23}^{-1}$&$+Z_{N}$      & $1$ & $-3$ &$2$&$0$&$\sqrt{15}$\\
\hline
\end{tabular}}}
\caption{$c_i$  and $\alpha$-$\beta$ coordinates  for $E_6$-motivated  $Z'$ models appearing in $E_6$ breakings.
We determine the $\pm$ signs in front  of  $Q_{mn}^l=-Q_{-m-n}^{-l}$  from the $\alpha$-$\beta$ parameterization in
Eq.~(\ref{eq:e6models}) and from Table~\ref{tab:ci}.  Models with the same multiplet structure appear in the same panel.}
}

\TABLE{\label{tab:models11}
\noindent\makebox[\textwidth]{
\scalebox{0.88}{
\begin{tabular}{|l|l|l|}
\hline
$U(1)^\prime$ & $Q_{mn}^{l}$ &\hspace{3cm}$ Q_{mn}^{l}$ \\ \hline
$U(1)_{R}$ & $+Q_{R00}^1$ & $(\eplus,\dbar,\lbar)_{+1} +(l,q,D,\ddbar,S)_{0} + (\nubar,\ubar,L)_{-1}$ \\
$U(1)_{I}$ & $-Q_{I00}^1$ & $(\nubar,\ddbar,L)_{+1} +(\lbar,q,\ubar,D,\eplus)_{0}+(S,\dbar,l)_{-1}$ \\
$U(1)_{A}$ & $+Q_{A00}^1$ & $(S,\ubar,l)_{+1} + (L,q,\dbar,D,\nubar)_{0}+(\eplus,\ddbar,\lbar)_{-1}$ \\ \hline
$U(1)_{33}$ & $+Q_{X-10}$  &  $(l,\lbar,L)_{-1} +(\ubar,\dbar,\ddbar)_{0} + (\eplus,\nubar,S)_{+2}+q _{+1}+D_{-2} $ \\
$U(1)_{21\ovl{R}}$ & $+Q_{R0-1}$  &  $(\eplus,\nubar,\lbar,L)_{-1}+ (q,D)_{0} +(l,S)_{+2} +(\ubar,\dbar)_{+1} + \ddbar_{-2}$ \\
$U(1)_{21\ovl{I}}$ & $+Q_{I0-1}$ &  $(S,\nubar,l,L)_{-1} + (q,D)_{0} +(\lbar,\eplus) _{+2}+(\ddbar,\dbar)_{+1} + \ubar_{-2}$ \\
$U(1)_{21\ovl{A}}$ & $-Q_{A0-1}$ &  $(S,\eplus,l,\lbar)_{-1} + (q,D)_{0} +(L,\nubar)_{+2}+(\ddbar,\ubar)_{+1} + \dbar_{-2}$ \\ \hline
$U(1)_{31R}$ & $-Q_{R11}$ &  $(\lbar,L,S)_{0}+q_{+1} +(\ubar,\dbar)_{-1}+(\eplus,\nubar)_{+3}+l_{-3}+\ddbar_{+2}+  D_{-2}$ \\
$U(1)_{31I}$ & $-Q_{I11}$ &  $(l,L,\eplus)_{0} + q_{+1} +(\ddbar,\dbar)_{-1}+ (S,\nubar)_{+3}+\lbar_{-3}+\ubar_{+2} + D_{-2}$ \\
$U(1)_{31A}$ & $+Q_{A11}$ &  $(l,\lbar,\nubar)_{0} + q_{+1}+(\ddbar,\ubar)_{-1}  +(S,\eplus)_{+3}+L_{-3}+ \dbar_{+2}+D_{-2}$ \\ \hline
$U(1)_{42R}$ & $+Q_{R1-1}$  &  $(\lbar,L,\ddbar,D)_{-2} +(\eplus,\nubar,l,q,\dbar,\ubar)_{+1} +  S_{+4}$ \\
$U(1)_{42I}$ & $-Q_{I1-1}$ &  $(l,L,\ubar,D)_{-2} + (S,\nubar,\lbar,q,\dbar,\ddbar)_{+1} + \eplus_{+4}$ \\
$U(1)_{42A}$ & $-Q_{A1-1}$ &  $(l,\lbar,\dbar,D)_{-2} +(S,\eplus,L,q,\ubar,\ddbar)_{+1} + \nubar_{+4}$ \\ \hline
$U(1)_{32R}$ & $+Q_{R1-2}$ &  $(\eplus,\nubar)_{0}+q_{+1} +(\ubar,\dbar)_{+2} + D_{-2}+l_{+3} + (\lbar,L)_{-3}+\ddbar_{-4}+S_{+6}$ \\
$U(1)_{32I}$ & $+Q_{I1-2}$ &  $(\nubar,S)_{0}+q _{+1} +(\dbar,\ddbar)_{+2}+D_{-2}+\lbar_{+3}+(l,L)_{-3}  + \ubar_{-4}+ \eplus_{+6}$ \\
$U(1)_{32A}$ & $-Q_{A1-2}$ &  $(\eplus,S)_{0}+q_{+1} +(\ubar,\ddbar)_{+2}+D_{-2}+L_{+3} + (l,\lbar)_{-3}+\dbar_{-4}+ \nubar_{+6}$ \\
$U(1)_{32\ovl{R}}$ & $+Q_{R-21}$ &  $l_{0}+(\ubar,\dbar)_{+1} + q_{+2} +\ddbar_{-2}+ (\eplus,\nubar)_{+3}+(\lbar,L)_{-3}+D_{-4}+S_{+6}$ \\
$U(1)_{32\ovl{I}}$ & $+Q_{I-21}$ &  $\lbar_{0}+(\dbar,\ddbar)_{+1}+q_{+2}+\ubar_{-2} +(l,L)_{-3}+(\nubar,S)_{+3}+D_{-4} +\eplus_{+6}$ \\
$U(1)_{32\ovl{A}}$ & $+Q_{A-21}$ & $L_{0}+(\ubar,\ddbar)_{+1}+q_{+2} +\dbar_{-2}+(\eplus,S)_{+3}+(l,\lbar)_{-3}+D_{-4} + \nubar_{+6}$ \\ \hline
$U(1)_{51R}$ & $-Q_{R21}$ &  $(\ubar,\dbar,\lbar,L)_{+1}+(q,\ddbar,S)_{-2}+(l,D)_{+4} + (\eplus,\nubar)_{-5} $ \\
$U(1)_{51I}$ & $-Q_{I21}$ &  $(l,\dbar,L,\ddbar)_{+1} + (q,\eplus,\ubar)_{-2} +(\lbar,D)_{+4}+ (\nubar,S)_{-5}$ \\
$U(1)_{51A}$ & $-Q_{A21}$ &  $(l,\ubar,\lbar,\ddbar)_{+1} +(q,\nubar,\dbar)_{-2}+(L,D)_{+4} + (\eplus,S)_{-5}$ \\
$U(1)_{51\ovl{R}}$ & $-Q_{R12}$ &  $ (q,\lbar,L)_{+1} +(\ubar,\dbar,D,S)_{-2}+ (\eplus,\nubar,\ddbar)_{+4}+l_{-5} $ \\
$U(1)_{51\ovl{I}}$ & $-Q_{I12}$ &  $(l,q,L)_{+1} + (\eplus,\dbar,D,\ddbar)_{-2} + (\nubar,\ubar,S)_{+4}+\lbar_{-5}$ \\
$U(1)_{51\ovl{A}}$ & $+Q_{A12}$ &$(l, q, \lbar)_{+1} +(\nubar,\ubar,D,\ddbar)_{-2}+ (\eplus,\dbar,S)_{+4} + L_{-5}$ \\ \hline
$U(1)_{41IA}$ & $+Q_{R-2-3}^{1}$ &  $\lbar_0+(L,q)_{+1}+(S,\ubar)_{-1}+(\dbar,D)_{-2}+(\eplus,\ddbar)_{+3} + \nubar_{+4}+l_{-4} $ \\
$U(1)_{41AR}$ & $+Q_{I-2-3}^{1}$ &  $L_{0}+(l,q)_{+1}+  (\eplus,\dbar)_{-1}+(\ddbar,D)_{-2}  + (\nubar,\ubar)_{+3}+S_{+4}+\lbar_{-4}$ \\
$U(1)_{41RI}$ & $-Q_{A-2-3}^{1}$ &  $l_{0} + (\lbar,q)_{+1}+(\nubar,\ddbar)_{-1}+(\ubar,D)_{-2} +(S,\dbar)_{+3}+\eplus_{+4}+L_{-4}$ \\
$U(1)_{41AI}$ & $+Q_{R-2-3}^{-1}$&  $L_0 + (\lbar,q)_{+1}+(S,\dbar)_{-1}+(\ubar,D)_{-2}+(\nubar,\ddbar)_{+3}+  \eplus_{+4}+l_{-4}$ \\
$U(1)_{41RA}$ & $+Q_{I-2-3}^{-1}$ &  $l_{0} + (L,q)_{+1} +(\eplus,\ddbar)_{-1}+(\dbar,D)_{-2} +(S,\ubar)_{+3}+\nubar_{+4}+\lbar_{-4}$ \\
$U(1)_{41IR}$ & $+Q_{A-2-3}^{-1}$ &  $\lbar_0+(l,q)_{+1} +(\nubar,\ubar)_{-1}+(\ddbar,D)_{-2}+ (\eplus,\dbar)_{+3}+S_{4} + L_{-4}$ \\ \hline
$U(1)_{\chi RI}$ & $+Q_{A-23}^{1}$ &  $S_{0}+(\eplus,q,\ubar)_{+1}+(L,\ddbar)_{+2}+(\lbar,D)_{-2}+(l,\dbar)_{-3}+\nubar_{+5}$\\
$U(1)_{\chi AR}$ & $-Q_{I-23}^{1}$ &  $\nubar_{0}+(S,q,\ddbar)_{+1}+(\lbar,\dbar)_{+2}+(l,D)_{-2}+(L,\ubar)_{-3}+\eplus_{+5}$\\
$U(1)_{\chi IA}$ & $-Q_{R-23}^{1}$ &  $\eplus_{0}+(\nubar,q,\dbar)_{+1}+(l,\ubar,)_{+2}+(L,D)_{-2}+(\lbar,\ddbar)_{-3}+S_{+5}$\\
$U(1)_{\chi IR}$ & $+Q_{A-23}^{-1}$ &  $\eplus_{0}+(S,q,\ddbar)_{+1}+(L,\ubar)_{+2}+(l,D)_{-2}+(\lbar,\dbar)_{-3}+\nubar_{+5}$\\
$U(1)_{\chi RA}$ & $-Q_{I-23}^{-1}$ &  $S_{0}+(\nubar,	q,\dbar)_{+1}+(\lbar,\ddbar)_{+2}+(L,D)_{-2}+(l,\ubar)_{-3}+\eplus_{+5}$\\
$U(1)_{\chi AI}$ & $-Q_{R-23}^{-1}$ &  $\nubar_{0}+(\eplus,q,\ubar)_{+1}+(l,\dbar)_{+2}+(\lbar,D)_{-2}+(L,\ddbar)_{-3}+S_{+5}$\\ \hline
\end{tabular}}}
\caption{Charge assignment for  $E_6$-motivated  $Z'$ models~(up to a normalization) appearing in $E_6$ breakings.
We determine the $\pm$ signs in front  of  $Q_{mn}^l=-Q_{-m-n}^{-l}$ as in Table~\ref{tab:models10}. 
Models with the same multiplet structure appear in the same panel.}}

\section{$E_6$  Chains of Subgroups}
\label{sec:breakings}
\FIGURE[t]{\label{fig:Emax}
\includegraphics[scale=0.5]{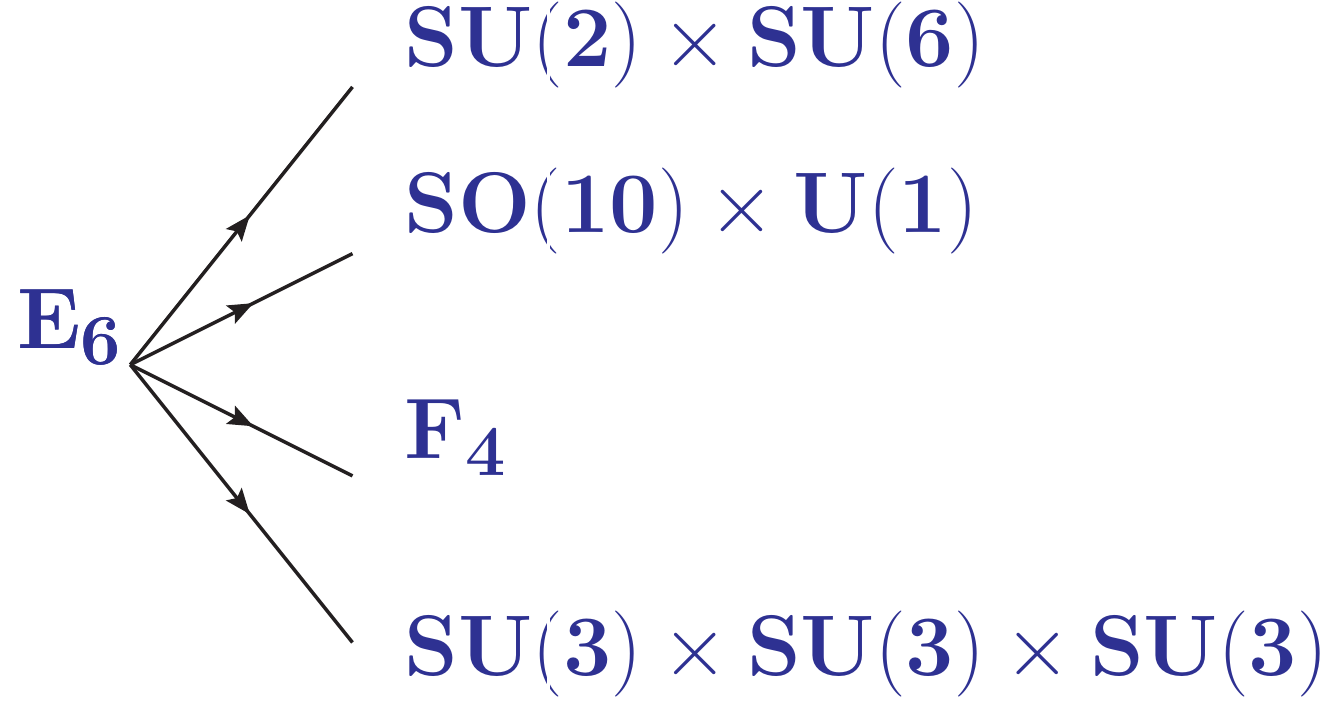}
\caption{$E_6 $ maximal subgroups.}}
All the $E_6$ breaking patterns have been considered in~\cite{Slansky:1981yr}
but there are different fermion assignments for the multiplets in a given breaking pattern. 
In the last section we studied how many different alternative models correspond to a given $U(1)'$.
Here we address the question whether these alternative models appear in chains of subgroups of $E_6$. 
As we will see, if a model appears in a known breaking pattern, 
then its alternative models will appear in the identical pattern (in most of the cases).
In this way, we find the set of all possible $U(1)'$ for a given breaking pattern. 
Once this is known, the orthogonality between the $Z'$ is enough to determine the $Z'$ models 
for every chain of  maximal subgroups.
In~\cite{Slansky:1981yr} the maximal subgroups of $E_6$ containing $U^{em}(1)\times SU(3)_C$ 
were shown to be $SU(2)\times SU(6)$, $SO(10)\times U(1)$, $F_4$ and $SU(3)\times SU(3)\times SU(3)$.
We now consider the subset of those cases containing the full SM group, $SU(3)_C\times SU(2)_L\times U(1)_Y$,
instead~(for a more detailed explanation see~\cite{Slansky:1981yr}). 

\subsection{$E_6\rightarrow SU(2)_X\times SU(6) $} 
Considering the first case in Figure~\ref{fig:Emax}, $SU(2)\rightarrow SU(2)_{X}$, with $X = R,I,A$, 
then for every chain of maximal subgroups all the $U(1)$ factors are uniquely defined by orthogonality (see Figure~\ref{fig:su2x}). 
This is because after breaking $SU(2)_X$ down to  $U(1)_X$,   
all other $U(1)'$ in this pattern should be perpendicular to $U(1)_X$.  
This constraint  is not present if we replace $SU(2)_X$ by the unbroken SM symmetry $SU(2)_L$.

\subsection{$E_6\rightarrow SU(2)_L\times SU(6) $}
By comparing Figure~\ref{fig:su2x} with Figure~\ref{fig:su2L}
corresponding to the breaking into $SU(2)_X\times SU(6)$ and $SU(2)_L\times SU(6)$, respectively,
the clearest difference appears in the further breaking into $SU(2)_L\times SU(5)\times U(1)_{51\bar{X}}$. 
The symmetry $U(1)_{51\bar{x}}$ is an alternative model for $U(1)_{51x}$ 
which allows  two possibilities for the the $SU(5)$ breaking, \ie
\begin{align}
SU(5)\rightarrow 
\begin{cases}
 SU(4)\times U(1)_{41XY} \\
 SU(3)\times SU(2)_X \times U(1)_{32 X}.
\end{cases}
\end{align}
Since $SU(2)_L$  is not broken,  there is just one constraint, namely the orthogonality to $U(1)_{51\bar{X}}$.
The models  $U(1)_{41XY}$~($Y\neq X$)  are perpendicular to $U(1)_{51\bar{X}}$
and $U(1)_{31\bar{y}}$~(see Figure~\ref{fig:su2L}) but they are not perpendicular to any $U(1)_X$. 
The difference between the two $SU(5)$ is that in one case $SU(2)_L \subset SU(5)$.

\FIGURE[t]{\label{fig:su2x}
\includegraphics[scale=0.7]{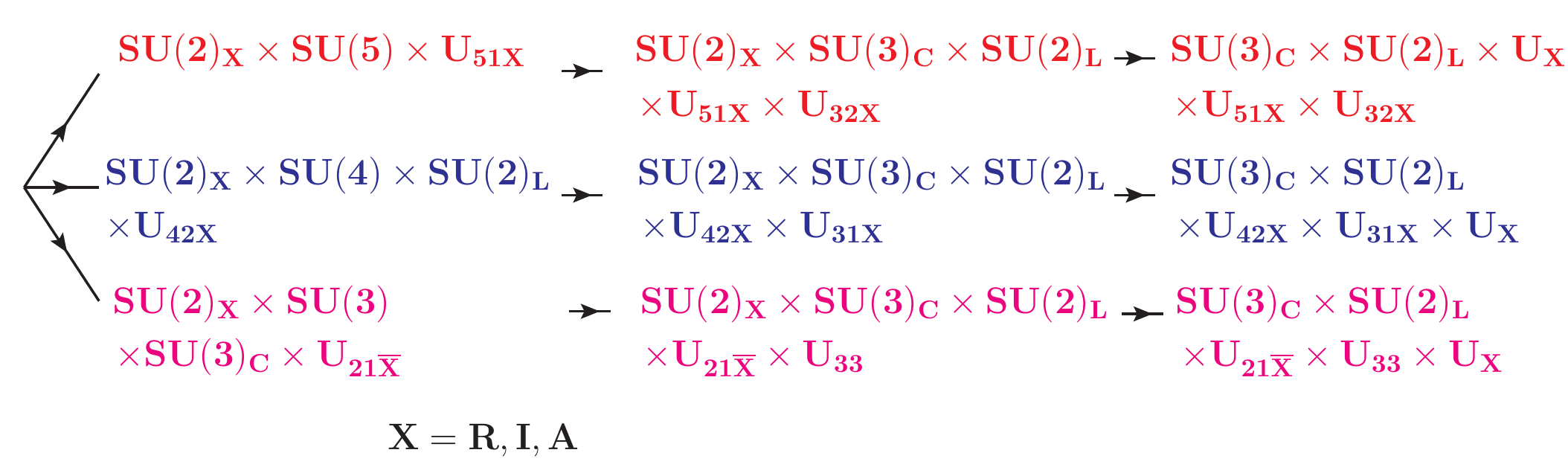}
\caption{$E_6 \rightarrow SU(2)_X\times SU(6)$ chains of subgroups.  All the $U(1)$ factors are uniquely defined for a fixed $X$. 
We recall our notation $U_{n-m\ m\ Z}\equiv U(1)_{n-m\ m\ Z}$.  The colors are used to distinguish between the different chains of
subgroups. }}

\subsection{$E_6\rightarrow U(1)\times SO(10)$}
The different fermion assignments for the breaking pattern $E_6\rightarrow U(1)\times SO(10)$
are displayed in Figure~\ref{fig:so10}. 
As shown in Table~\ref{tab:models1}, the model $Z_{\chi}$ corresponds to the $U(1)_{\chi RI}$
and has 5 alternative models, which are listed in Table~\ref{tab:models10}.
Figure~\ref{fig:so10} displays the chain of subgroups,
$U(1)_{42X}\times SU(5)\times U(1)_{\chi XY} \to U(1)_{42X}\times SU(3)_c\times SU(2)_L\times U(1)_{\chi XY}\times U(1)_{32Y}$,
which with the choice $X= R$ and $Y=I$ results in the ordinary $SU(5)$ unification group, 
with $U(1)_{\chi RI}$  and $U(1)_{32I}$  corresponding to the $Z_{\chi}$ and the hypercharge  $Z_Y$, respectively.   

The model  $Z_{N}$~\cite{Ma:1995xk,King:2005jy}  is associated with $U(1)_{\chi AI}$, and is an alternative model of the $Z_{\chi}$
appearing in the chain, $SO(10)\times U(1)_{42A} \rightarrow  SU(5)\times U(1)_{42A}\times U(1)_{\chi AI}$.
Similarly, for every model in Table~\ref{tab:models1}~(except  $Z_{S}$) we 
can find several chains of subgroups which contain them.
There exist two additional chains of subgroups which we do not show in Figure~\ref{fig:so10},
$$
E_6 \to U(1)_{42X}\times SO(10) \to U(1)_{42X}\times SO(9) \to U(1)_{42X}\times SO(7)\times U(1)_{X} \to 
$$
$$
U(1)_{42X}\times SU(4)\times U(1)_{X} \to U(1)_{42X}\times SU(3)_C\times U(1)_{X} \times U(1)_{31X}
$$
and the similar breaking pattern,
$$
E_6 \to U(1)_{42X}\times SO(10) \to U(1)_{42X}\times SO(9) \to U(1)_{42X}\times SU(4)\times U(1)_{X} \to 
$$
$$
U(1)_{42X}\times SU(3)_C\times U(1)_{X} \times U(1)_{31X}
$$
These patterns contain $U^{em}(1)\times SU(3)_C$, but not $SU(2)_L$, 
and therefore we do not consider them as options~(for further details see~\cite{Slansky:1981yr}).

\FIGURE[t]{\label{fig:su2L}
\includegraphics[scale=0.7]{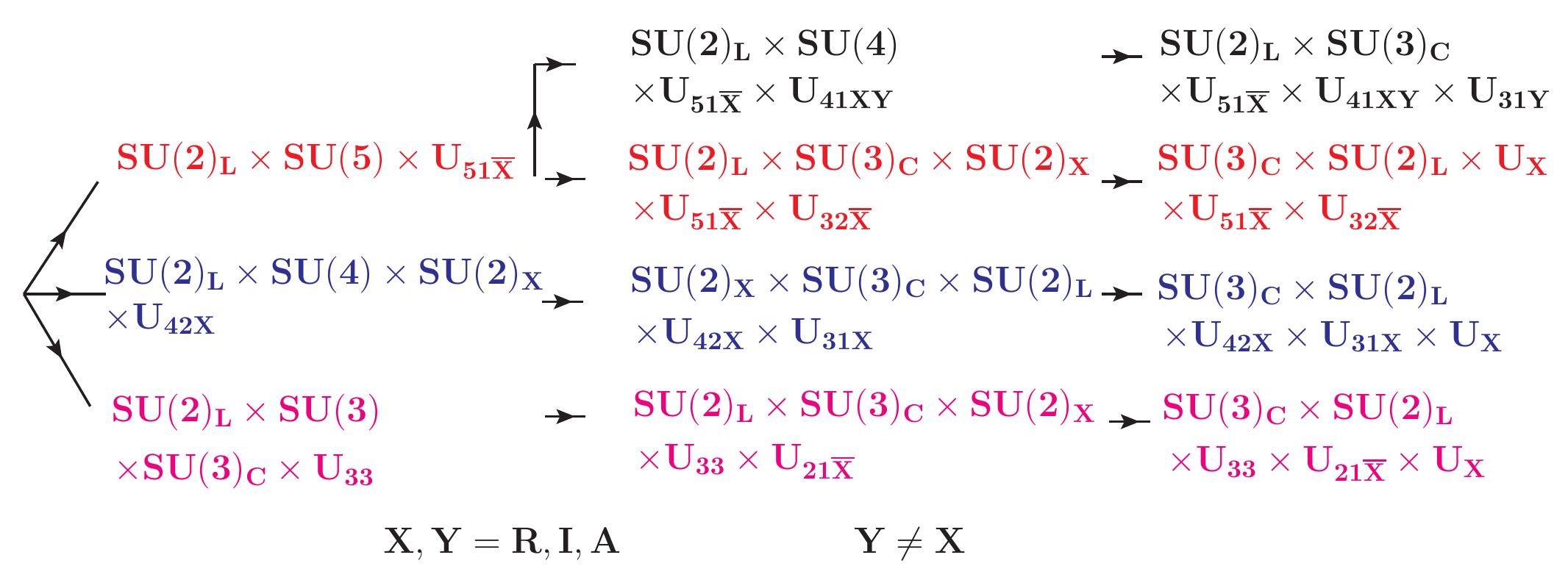}
\caption{Same as Figure~\ref{fig:su2x} but for $E_6 \rightarrow SU(2)_L\times SU(6)$ chains of subgroups. }}

\subsection{ $E_6\rightarrow SU(3)\times SU(3)\times SU(3)$ }

An important subgroup of $E_6$ for unified model building  is the ``trinification" group~\cite{Achiman:1979},
which has the same rank as $E_6$ and the dimension of its fundamental representation is 27 as in $E_6$. 
This subgroup appears in the chain  
$$
E_6\rightarrow SU(3)\times SU(3)\times SU(3)
\rightarrow SU(3)_C\times SU(2)_L\times U(1)_{33}\times SU(3)\rightarrow 
$$
$$
SU(3)_C\times SU(2)_L\times U(1)_{33}\times SU(2)_X\times U(1)_{21X} \rightarrow 
SU(3)_C\times SU(2)_L\times U(1)_{33}\times  U(1)_X   \times U(1)_{21X}
$$
Comparison with RR shows that  there are two additional models corresponding  to $X= I,A$.
For $X=I$ we find that the charges of the $U(1)_{I}$ do not contribute 
to electric charge~\cite{Witten:1985xc,London:1986dk}; 
thus, the diagonal generators of  $SU(3)_C\times SU(2)_L\times U(1)_{33}\times  U(1)_{21I}$
are enough to reproduce the electric charges of the fundamental 
representation of $E_6$.
The same holds for $SU(3)_C\times SU(3)_L\times  U(1)_{21I}$ which provides the basis for the class of 3-3-1 models\footnote{The relationship of these models 
with $E_6$ is explored in~\cite{Sanchez:2001ua}.}~\cite{Singer:1980sw,Pisano:1991ee,Frampton:1992wt,Montero:1992jk,Foot:1992rh}.

\subsection{$E_6\rightarrow F_4 \rightarrow  SO(9)$}
The  chains of subgroups starting with $E_6\rightarrow F_4 \rightarrow  SO(9)$ are similar 
to those containing $SO(9)$ in Figure~\ref{fig:so10},
the unique difference being the absence of the factor $U(1)_{42X}$. 
Due to the fact that  $F_4$  has  real or pseudo-real  
representations only\footnote{Other groups with only real or pseudo-real representations
include the orthogonal groups of odd dimension, the symplectic groups, $E_7$ and $E_8$.},
this kind of model predicts mirror fermions which have the same quantum numbers
with respect to the standard model group as the ordinary counterparts, quarks and leptons,
except that they have the opposite handedness~\cite{Maalampi:1988va}.
There are strong constraints on models predicting  this kind of fermions~\cite{Maalampi:1988va},
however they are satisfactory maximal subgroups in the sense 
that they contain $U_{\text{em}}(1)\times SU(3)_C$~(for further details and notation see \cite{Slansky:1981yr}).

\FIGURE[t]{\label{fig:so10}
\includegraphics[scale=0.7]{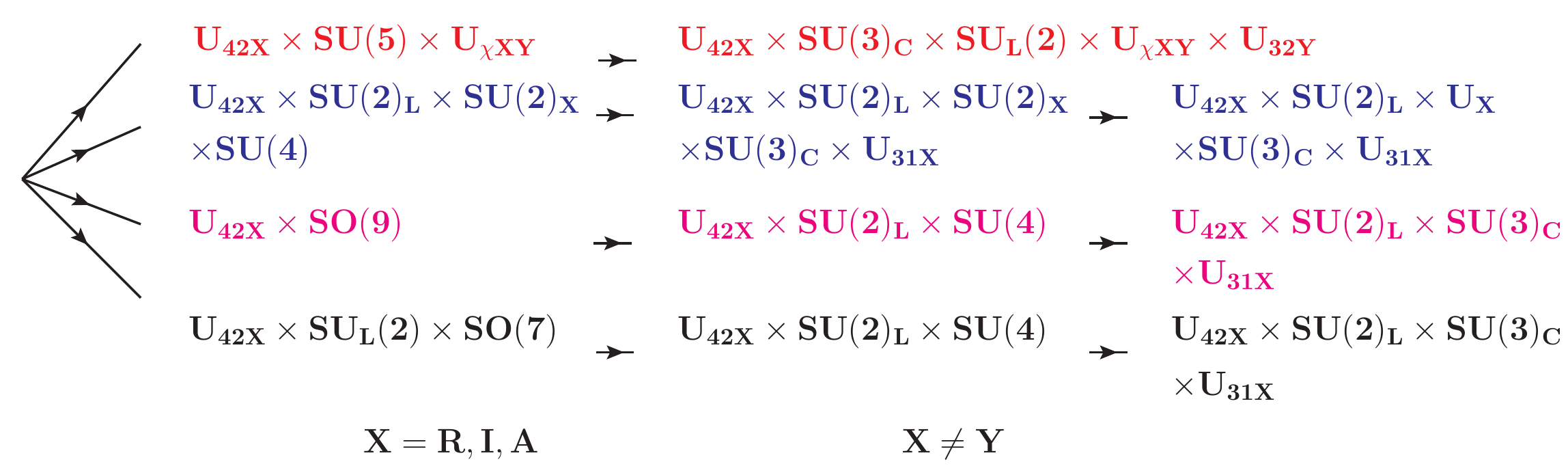}
\caption{Same as Figure~\ref{fig:su2x} but for $E_6 \rightarrow SO(10)\times U(1)_{42X}$ chains of subgroups. 
One can see from Table~\ref{tab:models10} that there two alternative models for the $Z_{\psi}$ and five for the $Z_{\chi}$.}}	

In summary, we have enumerated all the $E_6$ chains into maximal subgroups. 
The model charges and their coordinates appear in Tables~\ref{tab:models10} and \ref{tab:models11}.

\section{Low Energy Constraints on $E_6$}
\label{sec:low}

The effective parity-violating $e$-hadron and $e$-$e$ neutral-current interactions are
\begin{align}
 -\mathcal{L}^{eh}&=-\frac{G_f}{\sqrt{2}}\sum_{i}
\left[C_{1i}\ovl{e}\gamma_{\mu}\gamma^{5}e\ovl{q}_{i}\gamma^{\mu}q_{i}+
      C_{1i}\ovl{e}\gamma_{\mu}e\ovl{q}_{i}\gamma^{\mu}\gamma^{5}q_{i} \right],\\
-\mathcal{L}^{ee}&= -\frac{G_f}{\sqrt{2}}C_{2e}\ovl{e}\gamma^{\mu}\gamma^{5}e\ovl{e}\gamma^{\mu}e.
\end{align}
Setting  the $Z$--$Z'$ mixing angle equal to zero~\cite{Erler:2009jh,delAguila:2010mx}, 
and $\rho_{1} \equiv \frac{M_W^2}{M_Z^2 \cos \theta_W}=1$~(see~\cite{Langacker:1991pg}), 
then for $i=u,d$ we have
\begin{align}\label{eq:c1q}
C_{1i} = 2g_A^{1}(e)g_V^{1}(i)+2\rho_2 g_A^{\prime}(e)g_V^{\prime}(i),\notag\\
C_{2i} = 2g_V^{1}(e)g_A^{1}(i)+2\rho_2 g_V^{\prime}(e)g_A^{\prime}(i),\notag\\
\end{align}
where $\rho_2 \equiv (g^{\prime}M_Z)^2/(g_ZM_{Z'})^{2}$  and
\begin{align}
g_{V,A}^{1}(f)= \epsilon_L^1(f)\pm\epsilon_R^1(f),\hspace{1cm}
g_{V,A}^{\prime}(f)= \epsilon_L^2(f)\pm\epsilon_R^2(f),
\end{align}
are the corresponding vector and axial-vector couplings for the $Z$ and $Z'$ bosons. 
The quantities 
\begin{align}
 \epsilon_L^{1}(f)= T_{3}(f)-q(f)\sin^2 \theta_W^{\rm eff},\hspace{1cm} 
  \epsilon_R^{1}(f)= -q(f)\sin^2 \theta_W^{\rm eff},
\end{align}
are the effective couplings of the $Z$ boson to fermion $f$, where $T_{3}(f)$ and $q(f)$ are 
the third component of its weak isospin and its electric charge, respectively.
The low-energy effective mixing angle in the SM is 
$\sin^2 \theta^{\rm eff}_W = \kappa(0)\sin^2 \theta_W(M_Z)_{\ovl{MS}} = 0.23867$~\cite{Czarnecki:1995fw,Erler:2004in}. 
The chiral couplings for the $Z'$ are $\epsilon_L^2(f)=Q^{\prime}_L(f)$ and 
$\epsilon_R^2(f)=-Q^{\prime}_L(\ovl{f})$,  where the  $Q_{L}^{\prime}(f)$ are given for some models in Table~\ref{tab:models10}.

\FIGURE[t]{\label{fig:90alr}
\includegraphics[scale=0.5]{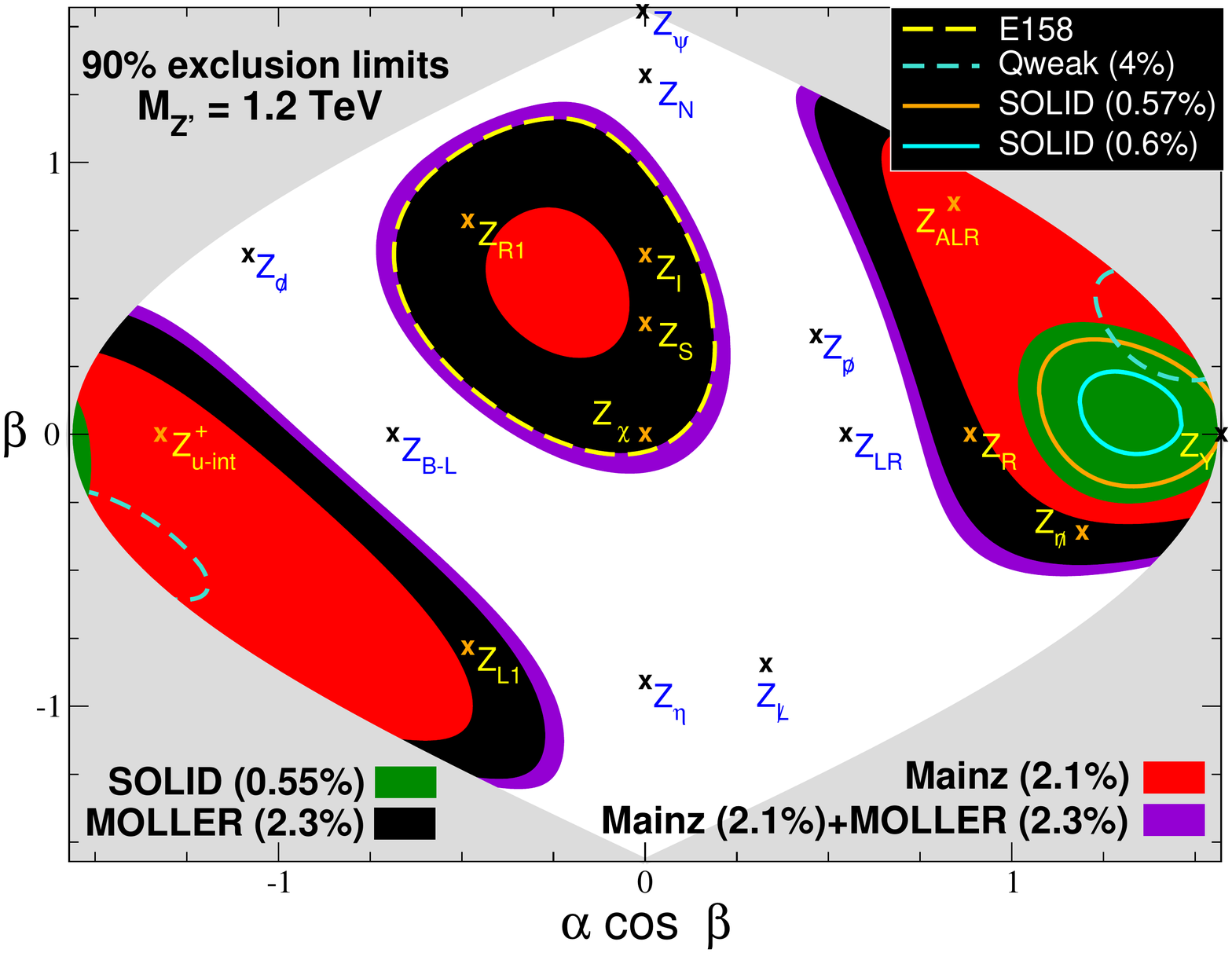}
\caption{$\alpha$-$\beta$ Sanson-Flamsteed projection of $E_{6}$ $Z'$ models.
The black, red and green colored regions correspond to the 90\% projected limits of
the MOLLER experiment with a relative precision of 2.3\%, the P2 Mainz  
proton weak charge measurement with a projected precision of 2.1\% and the SOLID experiment 
at JLab assuming a measurement of parity violation in deep inelastic scattering with a relative precision of 0.55\%. 
The yellow dashed contour encloses the 90\% excluded limit by E158. 
The continuous orange and cyan contours enclose the 90\% projected exclusion limits for a relative precision of 0.57\% and 0.6\%
in the measurement of $Q_{\text{SOLID}}$.  
For the projected limits we assume that no deviation of the SM expectation will be found in the planned experiments.}}

The scattering of polarized~(left or right-handed) electrons on an unpolarized target
allows the measurement of the left-right scattering asymmetry
\begin{align}\label{eq:alr}
 A_{LR}=\frac{d\sigma_L-d\sigma_R}{d\sigma_L+d\sigma_R}\ ,
\end{align}
where $d\sigma_{L,R}\equiv d\sigma(e^{-}_{L,R}e^{-}\rightarrow e^{-}_{L,R}e^{-})/dQ^2  $ is the differential cross-section in 
 the momentum transfer $Q^2$.  $A_{LR}$ differs from zero in the SM and at tree level it corresponds to
a measurement of the interference  between the $Z$ boson and the photon. 
The $A_{LR}$ asymmetry has been measured at low $Q^2 = 0.026$ GeV$^2$ in the SLAC--E158 experiment~\cite{Anthony:2005pm},
with the result 
$$A_{LR}=(1.31\pm 0.14~(\text{stat}.) \pm 0.10~(\text{syst}.))\times 10^{-7},$$
leading to a determination of the weak mixing angle of $\sin^2 \theta_W^{\rm eff} = 0.2403 \pm 0.0013$~\cite{Nakamura:2010zzi},
which is $1.25\sigma$ higher than the SM prediction~\cite{Czarnecki:1995fw,Erler:2004in}, $\sin^2 \theta^{\rm eff}_W =  0.23867$.
In the presence of a $Z'$ boson the relative change of $A_{LR}$ with respect to the SM expectation is given by~\cite{Li:2009xh,Chang:2009yw} 
\begin{align}\label{eq:moller}
\frac{A_{LR}-A_{LR}^{SM}}{A_{LR}^{SM}} = \frac{1}{\sqrt{2}G_F M_{Z'}^2}
\frac{g^{\prime 2}g^{\prime}_{V}(e)g^{\prime}_{A}(e)}{1-4\kappa(0)\sin^2 \theta_W(M_Z)_{\ovl{MS}}+\cdots},
\end{align}
where the dots stand for the one loop corrections given in~\cite{Czarnecki:1995fw}, $A_{LR}^{SM}$ is the 
expected value of $A_{LR}$ in the SM, $G_F$ is the Fermi constant
and $g^{\prime}=0.46151$~\cite{Erler:1999ug}. If we denote $\delta Q_W(e)$ as the change of the weak charge of the electron due 
to a $Z'$ then Eq.~(\ref{eq:moller}) is equal to $\delta Q_W(e)/Q_W(e)$, where $Q_W(e) =-2C_{2e}$ is the weak charge of the electron
in the SM ({\em cf.} Eq.~(\ref{eq:c1q})). 
With the upgraded electron beam at the Jefferson Laboratory~(Jlab) to 12~GeV 
a new project called MOLLER~(Measurement of Lepton-Lepton Electroweak Reaction) will improve the  E158 measurement of $Q_W(e)$
by a factor of 5~\cite{Kumar:2010zzf,Erler:2011iw}~(see Figures~\ref{fig:90alr} and \ref{fig:90alrh}). 
In Figures~\ref{fig:90alr} and \ref{fig:90alrh} the $Z_{\text{SOLID}}=Q_{R-10}^{1}$ is a boson with  vector couplings to the electron and 
the down quark and axial coupling to the up quark. Its  coordinates are  $\alpha = 0$ and  $\tan \beta = -\sqrt{3/5}$, 
and $c_1= 1$, $c_2= 0$,  $c_3= -1$.

\FIGURE[t]{
\label{fig:90alrh}
\includegraphics[scale=0.45]{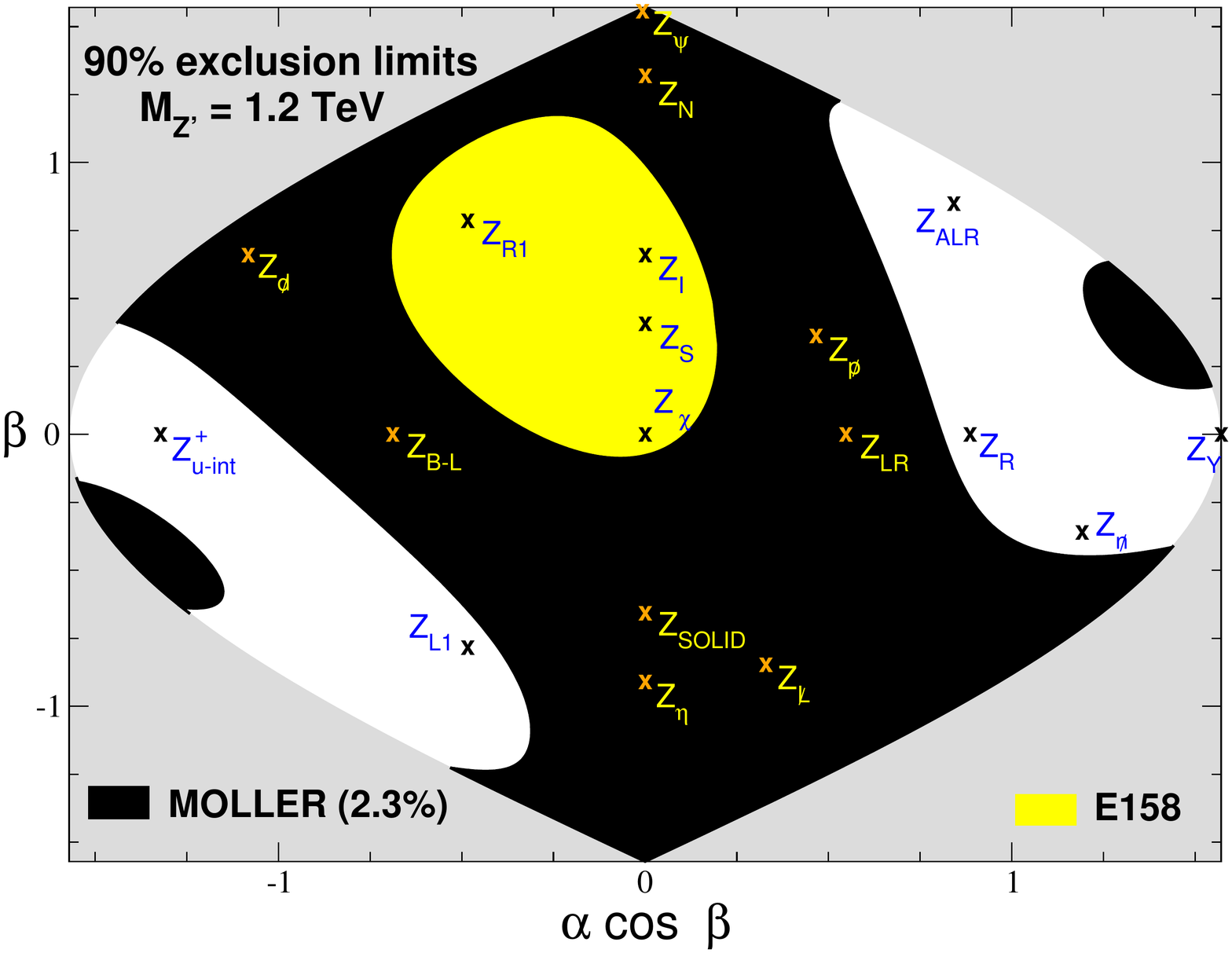}
\caption{$\alpha$-$\beta$ Sanson-Flamsteed projection of $E_{6}$ Z' models.
The yellow region corresponds to the 90\% exclusion limit from E158. 
The black region corresponds to the 90\% projected exclusion limit from MOLLER
for a precision of 2.3\%. In this case we assume a deviation in the measurement 
of $A_{LR}$ equal to half of the deviation of E158.}}

The Qweak experiment at JLab~\cite{vanOers:2007if,Androic:2013rhu} will be able to measure the weak charge of the
proton, $Q_W(p) = -2 [2 C_{1u} + C_{1d} ]$ and $\sin^2\theta_W^{\rm eff}$ in polarized $ep$ scattering with relative
precisions of $4\%$ and $0.3\%$, respectively~(see Figure~\ref{fig:90alr}). 
A similar experiment at the medium-energy accelerator MESA in Mainz, 
may be able to improve the precisions by a further factor of 2 or 2.5.
A very precise determination of the weak charge of $^{12} C$ may also be possible~\cite{Erler:2014fqa}.

The upgrade at Jlab will also allow precision measurements in parity-violating deep inelastic scattering. 
This project, known as SOLID~(Solenoidal Large Intensity Device)~\cite{Souder:2008zz,Kumar:2013yoa,Erler:2013xha,Souder}, would allow  $0.6\%$ measurements of $A_{LR}$~(see Figure~\ref{fig:90alr}).
One of the main goals of this experiment is the isolation of the linear combination $2C_{2u}-C_{2d}$, 
which is difficult to measure using  elastic scattering~\cite{Wang:2014bba,Wang:2014guo}. The left-right asymmetry in SOLID is proportional 
to $(2C_{1u} - C_{1d}) + 0.84(2C_{2u} - C_{2d})$.

The weak charge for an atom with $N$ neutrons and $Z$ protons is defined by
\begin{align}
Q_{W}(Z,N)=-2[C_{1u}(2Z+N)+C_{1d}(Z+2N)].
\end{align}
In the SM, $Q_{W}(Z,N)\approx Z(1-4\sin^2 \theta_W)-N\approx -N$.
There are precise experiments measuring atomic parity violation (APV) 
in cesium~(at the  0.4\% level~\cite{Wood:1997zq}) and other heavy atoms.

These experiments (will) provide very precise
determinations of the weak mixing angle off the $Z$ peak and will be sensitive to various
types of new physics~\cite{Erler:2014fqa,Kumar:2013yoa,Erler:2013xha,Erler:2004cx,Buckley:2012tc,GonzalezAlonso:2012jb}.

\section{Conclusions}
We have classified the two-dimensional $E_6$ parameter space of $U(1)$ symmetries by means of a systematic notation. 
This classification allows to identify $Z'$ models with the same multiplet structure and is convenient  
to determine the $U(1)$ factors for chains of maximal subgroups of $E_6$ and its alternative versions.
For these $U(1)$ groups we presented the $\alpha\text{-}\beta$ coordinates and the respective charges of the fundamental representation of $E_6$. 
We also used  low energy constraints from current and future parity violating asymmetry measurements and atomic parity non-conservation
in order to set 90$\%$ C.L. projected limits on the entire $E_6$ parameter space for a reference mass of $M_{Z'}=$ 1.2~TeV.  

\section*{Acknowledgments} 
It is a pleasure to thank Krishna Kumar for valuable suggestions regarding figures. 
J.E. is supported by PAPIIT (DGAPA--UNAM) project IN106913 and CONACyT (M\'exico) project 151234. 
E.R. acknowledges financial support provided  by FAPESP,  
``Patrimonio Aut\'onomo Fondo Nacional de Financiamiento para la Ciencia, la Tecnolog\'ia y la Innovaci\'on, Francisco Jos\'e de Caldas'' 
and ``Sostenibilidad-UDEA 2014--2015''. 

\bibliographystyle{apsrev4-1}
\bibliography{myreferences2}
\end{document}